\documentclass[aps,prl,preprint,nopacs,superscriptaddress]{revtex4-1}

\usepackage[utf8x]{inputenc}
\usepackage{amsmath}
\usepackage{amssymb}
\usepackage{graphicx}% Include  files
\usepackage{dcolumn}% Align table columns on decimal point
\usepackage{bm}% bold math
\usepackage{subfigure}
\usepackage{amsmath}

\usepackage{color}
\usepackage{colortbl}
\usepackage{ulem}
\usepackage{multirow}
\definecolor{mygray}{rgb}{1,1,0.9}
\definecolor{mycay}{rgb} {0.8,0.9,1} 
\makeatletter

    \def\CT@@do@color{%
      \global\let\CT@do@color\relax
            \@tempdima\wd\z@
            \advance\@tempdima\@tempdimb
            \advance\@tempdima\@tempdimc
    \advance\@tempdimb\tabcolsep
    \advance\@tempdimc\tabcolsep
    \advance\@tempdima2\tabcolsep
            \kern-\@tempdimb
            \leaders\vrule
                    \hskip\@tempdima\@plus  1fill
            \kern-\@tempdimc
            \hskip-\wd\z@ \@plus -1fill }
    \makeatother

\begin{document}

%\title{Nonsymmorphic symmetries breaking induced topological bulk-boundary distinctions} 
%\title{Surface symmetries breaking induced bulk-boundary distinctions \\ in nonsymmorphic topological chiral crystals} 
%\title{Bulk-boundary distinctions of nonsymmorphic topological chiral crystals} 
\title{Topological photocurrent responses from chiral surface Fermi arcs} 
%\title{Disentangled topological responses:\\ Photocurrents from surface Fermi arcs and bulk Weyl fermions} 
%\title{Surface photocurrents  from topological Fermi arcs} 
\author{Guoqing Chang}
\affiliation {Laboratory for Topological Quantum Matter and Advanced Spectroscopy (B7), Department of Physics, Princeton University, Princeton, New Jersey 08544, USA}

\author{Jiaxin Yin}\affiliation {Laboratory for Topological Quantum Matter and Advanced Spectroscopy (B7), Department of Physics, Princeton University, Princeton, New Jersey 08544, USA}
\author{Titus~Neupert}
\affiliation{Department of Physics, University of Zurich, Winterthurerstrasse 190, 8057 Zurich, Switzerland}

\author{Daniel~S.~Sanchez}\affiliation {Laboratory for Topological Quantum Matter and Advanced Spectroscopy (B7), Department of Physics, Princeton University, Princeton, New Jersey 08544, USA}

\author{Ilya Belopolski}\affiliation {Laboratory for Topological Quantum Matter and Advanced Spectroscopy (B7), Department of Physics, Princeton University, Princeton, New Jersey 08544, USA}

\author{Songtian S. Zhang}\affiliation {Laboratory for Topological Quantum Matter and Advanced Spectroscopy (B7), Department of Physics, Princeton University, Princeton, New Jersey 08544, USA}

\author{Tyler A. Cochran}
\affiliation {Laboratory for Topological Quantum Matter and Advanced Spectroscopy (B7), Department of Physics, Princeton University, Princeton, New Jersey 08544, USA}

\author{Ming-Chien Hsu }
\affiliation{Department of Physics, National Sun Yat-Sen University, Kaohsiung 804, Taiwan}

\author{Shin-Ming Huang}
\affiliation{Department of Physics, National Sun Yat-Sen University, Kaohsiung 804, Taiwan}

\author{Biao Lian}
\affiliation{Department of Physics, Princeton University, Princeton, New Jersey 08544, USA}

\author{Su-Yang Xu}
\affiliation {Laboratory for Topological Quantum Matter and Advanced Spectroscopy (B7), Department of Physics, Princeton University, Princeton, New Jersey 08544, USA}

\author{Hsin Lin}
\affiliation{Institute of Physics, Academia Sinica, Taipei 11529, Taiwan}

\author{M. Zahid Hasan$^{\dag}$\footnote[0]{$^{\dag}$Corresponding author (email):mzhasan@princeton.edu }} 
\affiliation {Laboratory for Topological Quantum Matter and Advanced Spectroscopy (B7), Department of Physics, Princeton University, Princeton, New Jersey 08544, USA}
\affiliation{Materials Sciences Division, Lawrence Berkeley National Laboratory, Berkeley, CA 94720, USA}


\date{\today}

\begin{abstract}

%In non-magnetic chiral crystals, real-space structural chirality gives rise to momentum-space Weyl fermions which further induce many exotic quantum properties, including giant chiral Fermi arc surface states and the  Weyl circular photogalvanic effect (CPGE) where a circularly-polarized laser generates a current from bulk Weyl fermions. Here we propose a new (hitherto unknown) surface-only  CPGE from chiral Fermi arc surface states. Using the ideal topological chiral crystal RhSi as a representative,  we quantitatively compute topological photocurrents from the giant chiral Fermi arcs. By rigorous crystal symmetry analysis, we demonstrate that Fermi arc photocurrents can be perpendicular to the bulk injection currents regardless of the choice of materials' surface. We then generalize this finding to all cubic chiral space groups and predict material candidates. Our theory reveals a powerful notion where common crystalline-symmetry breaking can be used to induce universal topological responses as well as making it possible to completely disentangle bulk and surface topological responses in many conducting material families.

The nonlinear optical responses from topological semimetals are crucial in both understanding the fundamental properties of quantum materials  and designing next-generation light-sensors or solar-cells. However, previous work was focusing on the optical effects from bulk states only,  disregarding topological surface responses. Here we propose a new (hitherto unknown) surface-only  topological photocurrent response from chiral Fermi arcs. Using the ideal topological chiral semimetal RhSi as a representative,  we quantitatively compute the topologically robust photocurrents from Fermi arcs on different surfaces. By rigorous crystal symmetry analysis, we demonstrate that Fermi arc photocurrents can be perpendicular to the bulk injection currents regardless of the choice of materials' surface. We then generalize this finding to all cubic chiral space groups and predict material candidates. Our theory reveals a powerful notion where common crystalline-symmetry can be used to induce universal topological responses as well as making it possible to completely disentangle bulk and surface topological responses in many conducting material families.

 \end{abstract}

\maketitle

Crystalline symmetries play a fundamental rule in determining the electronic and optical properties of topological materials \cite{Hasan2010, Qi2011, Weyl2018,Bansil2016,Burkov2016}. Materials in the same space groups may exhibit similar quantum properties due to the crystal symmetries they share \cite{TCI,NS-3DDirac,fillingconstraint,Nodalchain,Hourglass,DoubleDirac,NewFermion,Hopflink,KramersWeyl,Highorder}. By analyzing symmetry properties common to different space groups, one can generalize universal topological characteristics present across many material classes. For example, in non-magnetic chiral crystals, real-space structural chirality robustly gives rise to largely-separated Weyl-like fermions with quantized Chern numbers in energy-momentum space which further induced many emergent quantum properties ~\cite{KramersWeyl}, including exotic bulk circular photogalvanic effect (CPGE) where a circularly-polarized laser generates a current from Weyl fermions \cite{TaAsphoto_theory, TaAsphoto_exp, RhSi, RhSiphoto, QCPGE,RhSiphoto_exp,MoTephoto_exp,TaAs_exp2} and the giant Fermi arc surface states ~\cite{RhSi, CoSi, CoSi_exp, RhSi_exp,AlPt_exp,CoSi_exp2}. The large photocurrents and broad light-sensitive windows of topological chiral crystals with Weyl-like fermions can be used to realize next-generation light-sensors or solar-cells beyond conventional semiconductors. Due to the potential applications, the CPGE from Weyl-like semimetals has caught intense research interests in both theory and experiment. Both bulk Weyl fermions and surface Fermi arcs are important in Weyl-like semimetals. However, previous works only considered the photocurrents from bulk Weyl cones but overlooked the contributions from surface Fermi arcs \cite{KramersWeyl, TaAsphoto_theory, TaAsphoto_exp, RhSi, RhSiphoto, QCPGE,RhSiphoto_exp,MoTephoto_exp,TaAs_exp2}.

Here, we for the first time study a new CPGE from surface Fermi arcs, which is induced by the chiral structures of Fermi arcs and surface crystalline-symmetry breaking on the boundary. We first quantitatively compute the photocurrents from the giant Fermi arcs in the ideal Weyl-like semimetal RhSi and then generalize our theory to topological cubic chiral crystals in space groups \#195-199 and \#207-214. Regardless of the choice of surface terminations, the photocurrents from surface Fermi arcs can always be vertical to the injection currents from the bulk Weyl-like fermions in cubic chiral crystals.   In general, it is challenging and nontrivial to disentangle the contributions from bulk and surface in topological metals \cite{ARPES1, ARPES2, ARPES3, QOSC, CME_arc}. For example, in transport, the quantum oscillations from the bulk Dirac/Weyl cones are often mixed with those from the surface~\cite{ QOSC}.   Similarly, for spectroscopy, one needs to conduct multiple measurements to distinguish the bulk signals and surface states \cite{ARPES1}.  Our theory shows that different symmetry constraints in bulk and on the surface allow circumstances where the surface and bulk topological responses are completely disentangled.

\begin{figure*}
\includegraphics[width=158mm]{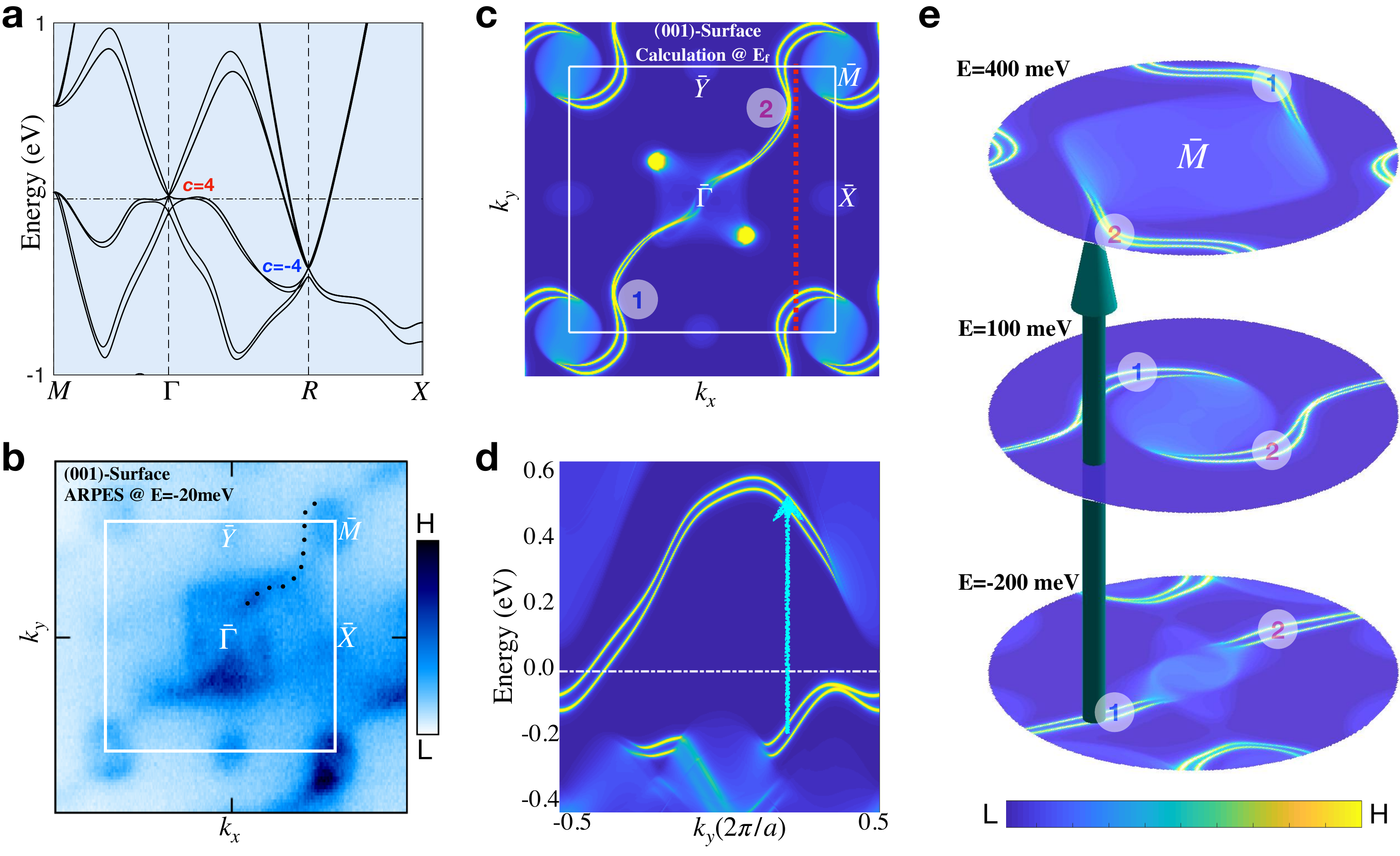}
\caption{\textbf{Spiral Fermi arcs on the (001)-surface of RhSi}.   \textbf{a,} Electronic structures of RhSi in the presence of spin-orbit coupling. \textbf{b,} ARPES-measured (001) surface states of RhSi at 20 meV below the Fermi level. The Fermi arc surface states (indicated by the black dot line) stretch across the entire surface Brillouin zone (BZ) along the $xy$-direction. The white box indicates the first BZ. The color-bar indicates the high (H) and low (L) carrier density distributions on the surface. \textbf{c,} First-principles calculations of (001) surface states of RhSi at the Fermi level. Our calculations match with the experimental measurements in panel $b$. \textbf{d,} The energy dispersion along the red dashed line indicated in panel $c$. The white dashed line indicates the Fermi level, and the cyan arrow shows the possible light-absorption channel between the Fermi arcs at different energies. \textbf{e,}  Helicoid Fermi arc structures around $\bar{M}$ at different energies. Fermi arcs spiral clockwise with increasing of energy,  which induce the potential light absorption channel indicated by the cyan arrow.}
\label{Fig1}
\end{figure*}

The RhSi family has a non-symmorphic cubic crystal structure in the space group $P2_{1}3$ (\# 198) with the 2-fold screw rotations $S_{2x}=\{C_{2x}|0.5,0.5,0\}$, $S_{2y}=\{C_{2y}|0,0.5,0.5\}$, and $S_{2z}=\{C_{2z}|0.5,0,0.5\}$, which are related by the 3-fold diagonal rotation $C_{3xyz}$. This material class has been predicted as ideal Weyl semimetals with a four-fold degenerate chiral fermion at the $\Gamma$ point and a six-fold degeneracy at the bulk  Brillouin zone (BZ) corner $R$ ~\cite{RhSi, CoSi}. The electronic structure of RhSi in the presence of spin-orbit coupling (SOC) is plotted in Fig. ~\ref{Fig1}a.   The angle-resolved photoemission spectroscopy (ARPES) measured constant energy contour on (001)-surface of RhSi is illustrated in Fig.~\ref{Fig1}b, where the bulk Weyl/chiral fermions projected at $\bar{\Gamma}$ (Chern number $C=4$ for the gap at the Fermi level)  and $\bar{M}$ ($C=-4$) are connected by Fermi arcs spanning diagonally across the entire surface BZ ~\cite{CoSi_exp, RhSi_exp,AlPt_exp,CoSi_exp2}. Figure ~\ref{Fig1}c shows the experimentally-matched surface states calculations of RhSi. In the absence of SOC, there are two sets of Fermi arc surface states. After turning on SOC, each set splits into two arcs of the opposite spins. The energy dispersion along the red path in Fig. ~\ref{Fig1}c cutting through the Fermi arcs is illustrated in Fig. ~\ref{Fig1}d. At one $k$-point there are two sets of Fermi arcs, which suggests a potential light-absorption channel (indicated by the cyan arrow) within Fermi arcs. This potential light-absorption channel is possible because of the chiral (or helicoid) structure of Fermi arc surface states \cite{RhSi_exp, Arc_Chen}. To better illustrate the chiral structure of Fermi arcs in RhSi, we plot the zoom-in of surface states around $\bar{M}$ in  Fig. ~\ref{Fig1}e. The Fermi arc surface states spiral clockwise with increasing energy.  As a result, the arc-set-1 and arc-set-2 can share the same momentum location at different energies.

This light-absorption channel of chiral Fermi arcs  suggests that the optical responses of the RhSi surface can be dramatically different from that in bulk. Here we study the  CPGE where a circularly polarized laser induces an injection current in the ideal Weyl/chiral semimetal RhSi.  Previous work was focusing on the photocurrents induced by the bulk Weyl nodes only, disregarding the CPGE contribution supported by the Fermi arc surface states ~\cite{TaAsphoto_theory, TaAsphoto_exp, RhSi, QCPGE,RhSiphoto_exp,MoTephoto_exp,TaAs_exp2}.  Recent experiments have just shown evidence the exotic bulk photocurrents in RhSi \cite{RhSiphoto_exp}. Therefore, it is timely to study the CPGE arising from giant Fermi arcs. In non-magnetic materials, the circularly polarized light induced injection current can be written as:

\begin{subequations}
\begin{eqnarray}
\frac{dJ_{i}}{dt} &=& \beta_{i,j}(\omega)[\textbf{E}(\omega) \times \textbf{E}(\omega)^{*}]_{j},
\label{eq:photocurrent1}
\\
 \beta_{i,j}(\omega)&=&\frac{\pi e^{3}}{\hbar^2 V} \epsilon_{jkl} \sum_{\textbf{k},n,m}  f_{nm}^{\textbf{k}} v_{\textbf{k},nm}^{i} r_{\textbf{k},nm}^{k} r_{\textbf{k},nm}^{l},
\label{eq:photocurrent2}
\end{eqnarray}
where $\textbf{E}(\omega)$ is the electric field of the laser and the subscript $j$ is the laser propagating direction, $i$ is the direction of injection current. Furthermore, $f_{nm}^{\textbf{k}} = f_{n}^{\textbf{k}}-f_{m}^{\textbf{k}}$ is the difference of Fermi-Dirac distributions between bands $n$ and $m$, $v_{\textbf{k},nm}^{i}=\frac{\partial E_{\textbf{k},nm}}{\partial k_{i}}$ is the difference of Fermi velocities, and $r_{\textbf{k},nm}^{l}=i\left\langle n|\partial H_{\textbf{k}}/\partial k_{l}|m\right\rangle$ is the matrix element of the velocity operator. $\beta$ is the CPGE tensor and $J$ is the injection current.  To distinguish the bulk and surface photocurrents, we use different superscripts: $\beta^{b}$ and $J^{b}$ for the bulk, and $\beta^{s}$ and $J^{s}$ for the surface.
\end{subequations}

We first study the CPGE of RhSi with the laser applied along the principal axes: $x$, $y$, or $z$ direction. For a generic point  $\textbf{k}=(k_{x}, k_{y}, k_{z})$ in bulk, the screw rotation $S_{2z}=\{C_{2z}|0.5,0,0.5\}$   relates $\textbf{k}$  to the partner $\textbf{k}'=(-k_{x}, -k_{y}, k_{z})$. The Fermi velocities and velocity matrix elements obey the relations $v_{\textbf{k}}^{x}=-v_{\textbf{k}'}^{x}$,  $v_{\textbf{k}}^{y}=-v_{\textbf{k}'}^{y}$, $v_{\textbf{k}}^{z}=v_{\textbf{k}'}^{z}$, $r_{\textbf{k}}^{x}=-r_{\textbf{k}'}^{x}$,  $r_{\textbf{k}}^{y}=-r_{\textbf{k}'}^{y}$, $r_{\textbf{k}}^{z}=r_{\textbf{k}'}^{z}$. From the integration of all the states in momentum space, we get $\beta_{x,z}^{\mathrm{b}}=0$ ($v_{\textbf{k}'}^{x}r_{\textbf{k}'}^{x}r_{\textbf{k}'}^{y}=-v_{\textbf{k}}^{x}r_{\textbf{k}}^{x}r_{\textbf{k}}^{y}$) and $\beta_{y,z}^{\mathrm{b}}=0$ ($v_{\textbf{k}'}^{y}r_{\textbf{k}'}^{x}r_{\textbf{k}'}^{y}=-v_{\textbf{k}}^{y}r_{\textbf{k}}^{x}r_{\textbf{k}}^{y}$). Owing to the cubic symmetry of the crystal, we further conclude that  $\beta_{y,x}^{\mathrm{b}}=0$,  $\beta_{z,x}^{\mathrm{b}}=0$,  $\beta_{x,y}^{\mathrm{b}}=0$, and  $\beta_{z,y}^{\mathrm{b}}=0$. The only nonzero bulk CPGE tensor elements are the diagonal terms $\beta_{x,x}^{\mathrm{b}}$, $\beta_{y,y}^{\mathrm{b}}$, and $\beta_{z,z}^{\mathrm{b}}$ which are parallel to the laser direction. We now look at the photocurrents of the Fermi arcs on the surfaces perpendicular to the laser. On the (001)-surface the screw rotation $S_{2z}$ is broken, since the spatial translation $\tau= (0.5, 0, 0.5)$ is no longer preserved on the boundary. Thus $\beta_{x,z}^{\mathrm{s}}$ and $\beta_{y,z}^{\mathrm{s}}$ from Fermi arcs are generically non-vanishing. Similarly, $\beta_{y,x}^{\mathrm{s}}$, $\beta_{z,x}^{\mathrm{s}}$, $\beta_{x,y}^{\mathrm{s}}$ and $\beta_{z,y}^{\mathrm{s}}$ can also non-zero. Therefore, on the (100)-, (010)-, and (001)-surfaces of RhSi, the Fermi arc surface states can generate photocurrents that are perpendicular to the laser direction.

\begin{figure*}
\includegraphics[width=158mm]{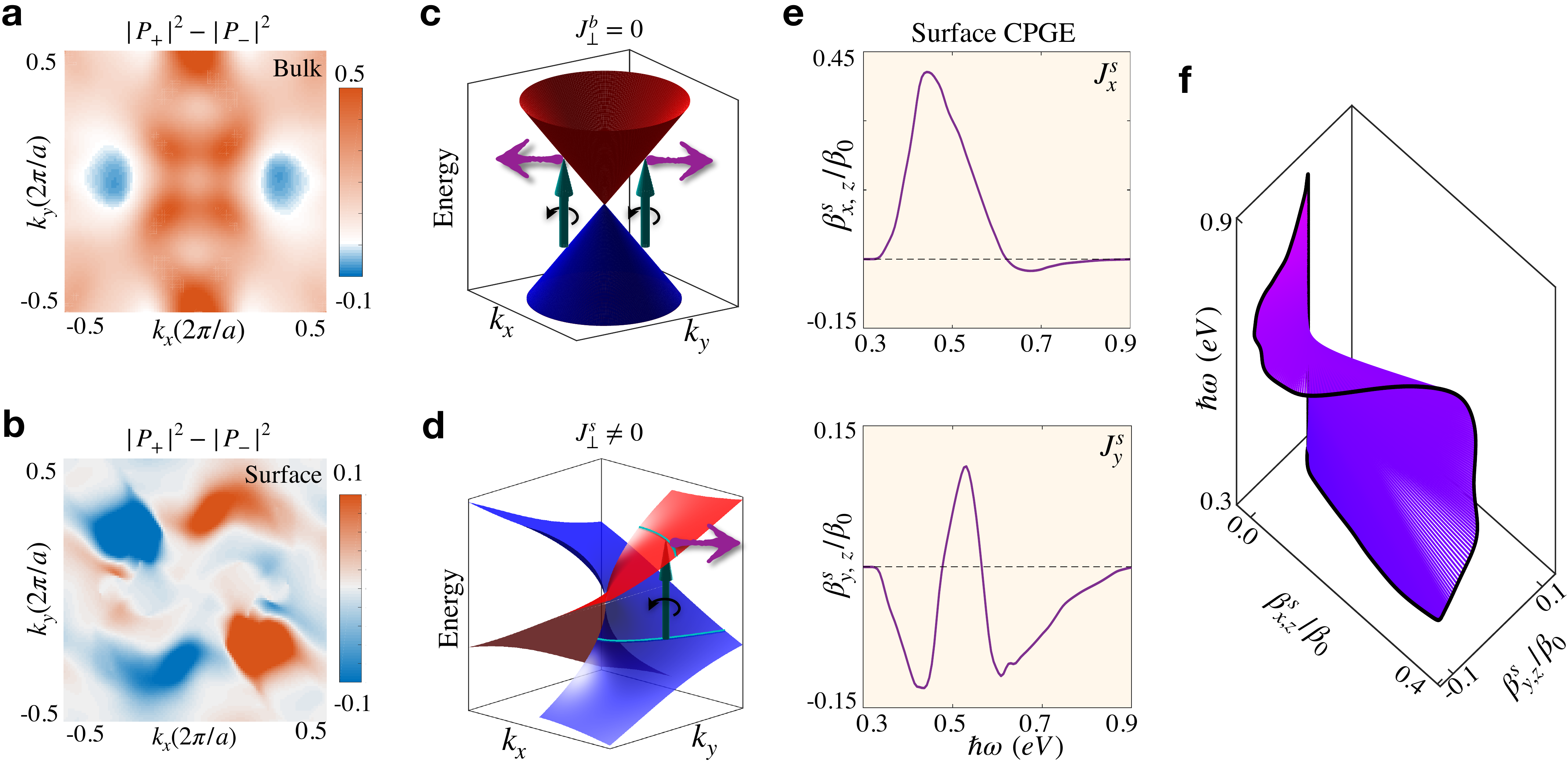}
\caption{\textbf{Circular photogalvanic effects (CPGE) in  RhSi with laser applied along $z$-axis.} \textbf{a,} The momentum-resolved circular light absorption of RhSi at $k_{z}=0.1\pi/a$ plane. The  $S_{2z}=\{C_{2z}|0.5,0,0.5\}$ rotation axis induces 2-fold light absorption from bulk cone. \textbf{b,} The light absorption of Fermi arc on the (001)-surface in RhSi. Only time-reversal symmetry is still preserved.  \textbf{c, }  Schematics of photocurrent in the $xy$ plane from the bulk Weyl cone. The cyan arrows indicate circularly polarized photons propagating along the $z$-direction, and the purple arrows indicate light-induced photocurrents. Due to the existence of  $S_{2z}$ rotation symmetry in bulk, the photocurrent perpendicular to the laser $J_{\perp}^{\mathrm{b}}$ vanishes. \textbf{d,}  Schematics of photocurrent from chiral Fermi arc surface states. The red and blue surface are two chiral Fermi arc surface states. A circularly polarized laser can only pump onside of chiral arc, and thus generate a net non-vanishing photocurrent. \textbf{e,} (001)-surface Fermi arcs induced photocurrents along $x$ and $y$ directions in RhSi,  respectively. $\beta_{0}=\pi e^{3}/h^{2}$, where $e$ and $h$ are the electron charge and the Planck constant.  \textbf{f,} The three-dimensional cartoon illustrates the direction and amplitude variations of the (001)-surface Fermi arcs induced photocurrent versus photon energy.}
\label{Fig2}
\end{figure*}

 The difference in the CPGE between the bulk and the (001)-surface under a circularly polarized laser along $z$-direction can be visualized by the momentum dependent circular optical absorption $|P_{+}|^{2}$-$|P_{-}|^{2}$ (Fig. ~\ref{Fig2}a).  Here, $P_{\pm}(\textbf{k})=1/\sqrt2[P_{x}(\textbf{k}) \pm iP_{y}(\textbf{k})]$ is the transition matrix element of circularly polarized light, and  $\textbf{P}(\textbf{k})=\left\langle\phi_{c}(\textbf{k})|\hat{\textbf{p}}|\phi_{v}(\textbf{k})\right\rangle$ is the interband transition \cite{MoS2}.  Under the circularly polarized laser, bulk electrons at \textbf{k} and the rotation partner $\textbf{k}'$ will be both  excited (Fig. ~\ref{Fig2}a) and generate opposite photocurrents J$_{\perp}^{\mathrm{b}}$ (photocurrents perpendicular to the laser) which then cancel with each other (Fig. ~\ref{Fig2}c). In contrast, on the surface of the crystal, electron excitations only occur on one side of the BZ (Fig.~\ref{Fig2}b) and thus produce net nonzero injection currents (Fig.~\ref{Fig2}d).  We quantitatively compute the photocurrents from the Fermi arcs (Fig.~\ref{Fig1}c) on the (001)-surface of RhSi. Our simulations are based on the Wannier functions derived from first-principles calculations. The photon energy dependent injection currents from the Fermi arc surface states of RhSi are plotted in Fig.~\ref{Fig2}e.  We observe a variation of both the direction and the amplitude of photocurrents of the Fermi arcs with the photon energy (Fig.~\ref{Fig2}f).

\begin{figure*}
\includegraphics[width=158mm]{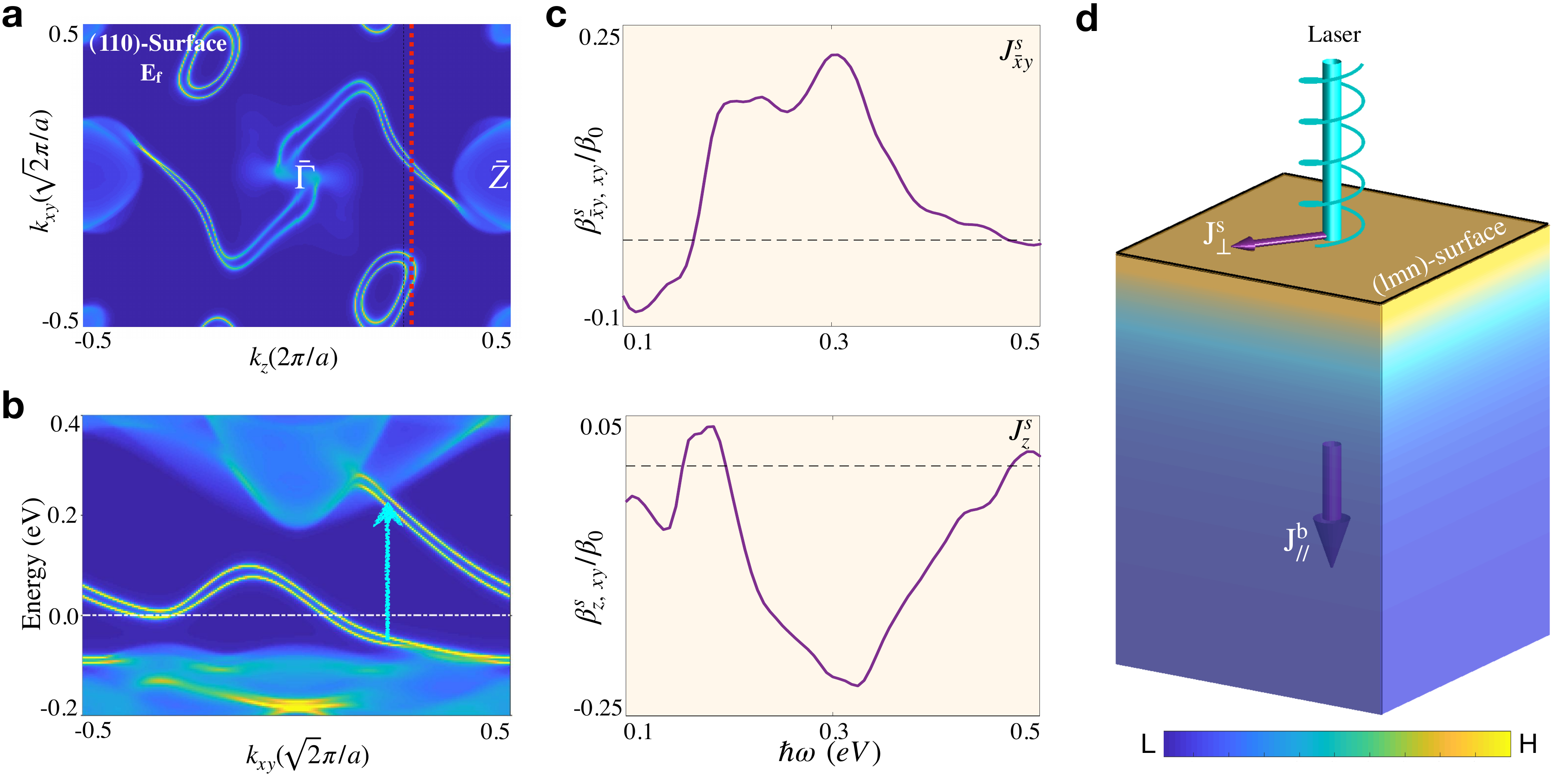}
\caption{\textbf{CPGE from Fermi arcs on generic surfaces of RhSi.} \textbf{a,}  (110)-surface states of RhSi at the Fermi level. Long Fermi arcs connecting projected chiral fermions at $\bar{\Gamma}$ and $\bar{Z}$ are observed.  \textbf{b,}  Energy dispersions cutting through the Fermi arcs on panel $a$. The cyan arrow indicates the possible light-absorption between the (110)-surface Fermi arc surface states. \textbf{c,} Numerical calculations reveal non-zero photocurrents from the (110)-surface Fermi arcs in RhSi under a laser vertical to the surface. \textbf{d,} 
For any generic (lmn)-surface of RhSi, under the circularly polarized laser vertical to the surface, the Fermi arcs induced photocurrents  ($J_{\bot}^{\mathrm{s}}$) are always perpendicular to the bulk photocurrents from chiral fermions ($J_{\parallel}^{\mathrm{b}}$).  The color-bar indicates the high (H) and low (L) carrier density distributions from Fermi arc surface states. The photocurrents from Fermi arcs should also be bounded on the surface, due to the surface localization of Fermi arcs.}
\label{Fig3}
\end{figure*}

Now we investigate the CPGE of RhSi when the laser is vertical to a generic (lmn)-surface. The laser along a generic direction can be decomposed under the three principal axes: $\textbf{R}=\frac{(l \hat{\textbf{x}},m \hat{\textbf{y}},n \hat{\textbf{z}})}{\sqrt{l^2+m^2+n^2}}$.  In bulk, if we only consider the $\hat{\textbf{x}}$ component of the electric field, the injection current is also along $x$-axis. And according to Eq.~\eqref{eq:photocurrent1}, the amplitude of photocurrents is proportional to the magnitude of the electric field. Same results are also applied to the photocurrent components along $y$- and $z$-directions. Therefore the bulk injection currents induced by the circularly polarized laser along $\textbf{R}$ direction can be written as: $\textbf{J}^{b}$=$|J_{\parallel}^{b}|(l \hat{\textbf{x}}+m \hat{\textbf{y}}+n \hat{\textbf{z}})$, which is parallel to the laser direction.  On  a generic (lmn)-surface where the rotation symmetries are  broken, surface photocurrents can be non-zero. We note that Fermi arcs surface states are always topologically protected.  Thus the in-plane injection currents from Fermi arcs which are perpendicular to the laser are allowed: $\textbf{J}^{s}_{\bot}$. Here we use the (110)-surface as the representative.  Figure. ~\ref{Fig3}a shows the isoenergetic contours of the (110)-surface, where long chiral Fermi arcs connect $\bar{\Gamma}$ and $\bar{Z}$. The energy dispersions of the chiral Fermi arc surface states are plotted in Fig. ~\ref{Fig3}b, with a light-absorption channel indicated by the cyan arrow. Figure ~\ref{Fig3}c shows the in-plane phtotcurrents of the (110)-surface of RhSi.   Figure ~\ref{Fig3}d illustrates the CPGE of RhSi when the circularly polarized laser is perpendicular to a generic (lmn)-surface. The photocurrents from chiral Fermi arc surface states, if any, are always perpendicular to the injection currents from the bulk cone. Therefore, by measuring photocurrents along different directions, one can detect contributions from bulk Weyl cones and surface Fermi arcs in RhSi separately.  Our theory provides a rare example where bulk and surface topological responses are fully disentangled.

Finally, we generalize our theory of disentangled photocurrents from Fermi arc surface states and bulk chiral fermions into other materials. In principle, all Weyl semimetals satisfying the following criteria can exhibit related photocurrents induced by chiral Fermi arcs: (1) Large nontrivial energy windows that allow an optical transition between two arcs at the same $\textbf{k}$-points; (2)  Low surface symmetries so that nonzero net surface photocurrents are allowed. In addition, in order to distinguish the photocurrents from Weyl cones and Fermi arcs, (3) additional crystal symmetries that eliminate bulk photocurrents perpendicular to the laser are required. Taking all the three criteria into consideration, we find that topological cubic chiral crystals in space groups \#195-199 and \#207-214 are the best material candidates.  The rotation symmetries of the cubic chiral crystals force the bulk photocurrents to be parallel to the laser. The unconventional multifold chiral fermions of cubic chiral crystals are largely separated in energy-momentum space, which prepare  long chiral Fermi arcs within a large energy window.

Table~\ref{symChiral} shows the CPGE of cubic chiral space groups with more material candidates for further experimental measurements. We first check the non-symorphic chiral space groups: \#198, \#199, \#208, \#210, and \#212-214.  By analyzing the symmetry constraints, we find that space groups \#198, \#212, and \#213 allow for the photocurrents from the (001)-surface, since the nons-ymmorphic symmetries are broken on the boundary. In contrast, in space groups \#199 and \#213 the CPGE from the Fermi arcs on the (001)-surface is disallowed since the spiral translations are preserved on the boundary. Although the screw rotation $S_{4z}$ is broken in space groups \#208 and \#210, the $C_{2z}$ rotation is still preserved. Therefore, the (001)-surface Fermi arcs induced photocurrents are also forbidden in space groups \#208 and \#210. For the other surface terminations where all the rotation symmetries are broken, the (110)-surface for example, photocurrents from Fermi arcs are allowed in all the non-symmorphic cubic chiral space groups. Lastly we study the Fermi arcs CPGE from symmorphic cubic chiral space groups: \#195-197, \#207, \#209, and \#211. On the (001)-surface, the rotation symmetries perpendicular to the plane is still preserved. Thus Fermi arcs induced photocurrents are forbidden. On other generic surfaces such as the (110)-surface, Fermi arc photocurrents are allowed due to the broken of rotation symmetry on the boundary. By extensive material search, we predict new materials (Table~\ref{symChiral}) which are ideal candidates to realize our theory.In general, the contributions of photocurrents from Fermi arcs can exist in all Weyl semimetals. However, in most materials, the photocurrents from surface Fermi arcs are mixed with the photocurrents from bulk, which makes it is challenging to distinguish the contributions from the surface and bulk. The material candidates in Table~\ref{symChiral} provide platforms for further experimental measurements of nonlinear optical responses from surface Fermi arcs.

\bigskip

To summarize, we have proposed new topological responses from surface Fermi arc surface states induced by the lifting of symmetry constraints on the surface. We have first computed the bulk and surface photocurrent contributions for the example of the RhSi class of materials from first-principles calculations, and then generalized our theory to all cubic chiral space groups with substantial material candidates to realize our proposal. We provide for the first time a comprehensive study of the hidden universal topological effects induced by crystalline-symmetry breaking in different space groups. Our theory provides a rare example where different symmetry constraints,  symmetry protection in bulk and crystalline-symmetry breaking on the surface, can be used to disentangle bulk and surface topological responses.

  \begin{center}
\begin{table*}

\begin{tabular}{|>{\columncolor{mycay}\sf}m{2.2cm}<{\centering}|>{\columncolor{mygray}\sf}m{2.6cm}<{\centering}|>{\columncolor{mygray}\sf}m{2.9cm} <{\centering}|>{\columncolor{mygray}\sf}m{2.9cm} <{\centering}|>{\columncolor{mygray}\sf}m{4.5cm}<{\centering}|}

\hline

\rowcolor{mycay}

{\bf{Space Group}} &\bf{Related Symmetries}&{\bf{Laser vertical to the (001)-surface }}&{\bf{Laser vertical to the (110)-surface}}&\bf{Material Candidates} \\
\hline
\hline
\hline
198 & $\{C_{2z}|\frac{1}{2},0,\frac{1}{2}\}$ & $J_{z}^{b}$;  $J_{x}^{s}$, $J_{y}^{s}$  & $J_{xy}^{b}$;  $J_{\bar{x}y}^{s}$, $J_{z}^{s}$ & AB (A=Co, Rh; B=Si, Ge); AuBe; AlX(X=Pd, Pt); BaPtY (Y=P, As) \\
\hline
199 & $\{C_{2z}|0,\frac{1}{2},0\}$ & $J_{z}^{b}$ &   $J_{xy}^{b}$;  $J_{\bar{x}y}^{s}$, $J_{z}^{s}$ & \\
\hline
208 & $C_{2z}$; $\{C_{4z}|\frac{1}{2},\frac{1}{2},\frac{1}{2}\}$   & $J_{z}^{b}$ & $J_{xy}^{b}$;  $J_{\bar{x}y}^{s}$, $J_{z}^{s}$  &  H$_{3}$X (X=P, As) \\
\hline
210 &  $C_{2z}$; $\{C_{4z}|\frac{1}{4},\frac{1}{4},\frac{1}{4}\}$  & $J_{z}^{b}$ &  $J_{xy}^{b}$;  $J_{\bar{x}y}^{s}$, $J_{z}^{s}$ & \\
\hline
212 & $\{C_{2z}|\frac{1}{2},0,\frac{1}{2}\}$; $\{C_{4z}|\frac{3}{4},\frac{1}{4},\frac{3}{4}\}$   & $J_{z}^{b}$;  $J_{x}^{s}$ $J_{y}^{s}$ & $J_{xy}^{b}$;  $J_{\bar{x}y}^{s}$, $J_{z}^{s}$  &  Li$_{2}$BX$_{3}$ (X=Pd, Pt)\\
\hline
213 & $\{C_{2z}|\frac{1}{2},0,\frac{1}{2}\}$; $\{C_{4z}|\frac{1}{4},\frac{3}{4},\frac{1}{4}\}$   & $J_{z}^{b}$; $J_{x}^{s}$, $J_{y}^{s}$ & $J_{xy}^{b}$;  $J_{\bar{x}y}^{s}$, $J_{z}^{s}$  & Mg$_{3}$Ru$_{2}$; Na$_{4}$Sn$_{3}$O$_{8}$; V$_{3}$Ga$_{2}$N; Nb$_{3}$Al$_{2}$N   \\
\hline
214 & $\{C_{2z}|0,\frac{1}{2},0\}$   & $J_{z}^{b}$ & $J_{xy}^{b}$;  $J_{\bar{x}y}^{s}$, $J_{z}^{s}$  &  La$_{3}$SiBr$_{3}$; La$_{3}$GaI$_{3}$\\
\hline
\hline
195-197 & $C_{2z}$ & $J_{z}^{b}$ &  $J_{xy}^{b}$;  $J_{\bar{x}y}^{s}$, $J_{z}^{s}$  & La$_{4}$Re$_{6}$O$_{19}$\\
\hline
207; 209;  211 & $C_{2z}$; $C_{4z}$ & $J_{z}^{b}$ &  $J_{xy}^{b}$;  $J_{\bar{x}y}^{s}$, $J_{z}^{s}$  & \\
\hline
\end{tabular}

\caption{ \textbf{Photocurrents of cubic chiral space groups and potential material candidates.}  Space groups \#198, \#199, \#208, \#210, and \#212-214 are nonsymmorphic chiral space groups, which include screw rotation axes indicated in the 2nd column. Space groups \#195-197, \#207, \#209, and \#211 are symmorphic chiral space groups.  Columns 3 and 4 show the nonvanishing photocurrents under circularly polarized laser along different directions. The electronic structures of the potential material candidates are shown in the Supplementary Section B.} 
\label{symChiral}
\end{table*}
\end{center}

\end{document}